\documentclass[reprint,final,floatfix,nofootinbib,superscriptaddress,longbibliography]{revtex4-2}

\usepackage{lmodern,mathtools,amssymb,mathrsfs,bbm,microtype,bm} 
\usepackage{color,graphicx} 
\definecolor{bluesmoke}{rgb}{0.207843,0.415686,0.623529}
\usepackage[
  allcolors=bluesmoke,
  colorlinks,
  pdftex,
  pdftitle={Phase transitions and microphases in elastomers.  I. Emergence of stable domains},
  pdfauthor={Manu Mannattil, Haim Diamant, and David Andelman},
  pdfkeywords={Phase separation, pattern formation, elasticity, elastomers, phase-field modeling},
  pdfsubject={Statistical Physics, Polymers \& Soft Matter, Condensed Matter \& Materials Physics},
  hypertexnames=false
]{hyperref}
\DeclareMathOperator{\sinc}{sinc}
\DeclareMathOperator{\trace}{tr}

\def\abs#1{\ensuremath{\left|#1\right|}}
\def\Am{\ensuremath{A_{\text{m}}}}
\def\dd{\ensuremath{\mathrm{d}}}
\def\div{\ensuremath{\boldsymbol{\nabla\cdot}}}

\def\grad{\ensuremath{\boldsymbol{\nabla}}}

\def\kbt{\ensuremath{k_{\text{B}}}T}
\def\mz{\ensuremath{m_{0}}}
\def\ms{\ensuremath{m_{\text{s}}}}
\def\NN{\ensuremath{\nonumber}}
\def\phic{\ensuremath{\phi_{\text{c}}}}
\def\phiz{\ensuremath{\phi_{0}}}
\def\psiz{\ensuremath{\psi_{0}}}
\def\qm{\ensuremath{q_{\text{m}}}}
\def\bqm{\ensuremath{\bm{q}_{\text{m}}}}
\def\strain{\ensuremath{\bm{\varepsilon}}}

\def\Tc{\ensuremath{T_{\text{c}}}}
\def\Tm{\ensuremath{T_{\text{m}}}}
\def\trans{\ensuremath{^{\mathsf{T}}}}

\def\gammaz{\ensuremath{\gamma_{0}}}
\def\unit{{\kern2.333pt}}

\definecolor{hlcolor}{rgb}{0.646,0.165,0.165}

\makeatletter
 \def\bibsection{%
  \par
  \begingroup
  \baselineskip26\p@
  \bib@device{\hsize}{72\p@}%
  \endgroup
  \nobreak\@nobreaktrue
  \addvspace{19\p@}%
}%
\makeatother

\begin{document}

\title{Phase transitions and microphases in elastomers.  I.~Emergence of stable domains}
\author{Manu Mannattil}
\email{manu.mannattil@posteo.net}
\thanks{Present address: School of Engineering and Applied Sciences, Harvard University, Cambridge, Massachusetts 02138, USA.}
\affiliation{School of Chemistry, and Center for Physics and Chemistry of Living Systems, Tel Aviv University, Tel Aviv 69978, Israel}
\affiliation{School of Physics and Astronomy, and Center for Physics and Chemistry of Living Systems, Tel Aviv University, Tel Aviv 69978, Israel}
\author{Haim Diamant}
\email{hdiamant@tauex.tau.ac.il}
\affiliation{School of Chemistry, and Center for Physics and Chemistry of Living Systems, Tel Aviv University, Tel Aviv 69978, Israel}
\author{David Andelman}
\email{andelman@tauex.tau.ac.il}
\affiliation{School of Physics and Astronomy, and Center for Physics and Chemistry of Living Systems, Tel Aviv University, Tel Aviv 69978, Israel}

\begin{abstract}
  Elasticity often plays a key role in regulating phase separation in physical systems.
  Recent experiments have shown that elastic effects can be used to control microphase separation in swollen elastomers.
  Here, microphase separation arises from a mismatch between the characteristic length scales of elastic and thermodynamic interactions.
  In this first part of a two-part paper, we show that microphase formation in elastomers
  can be explained using conventional theories of elasticity through a nonlocal thermodynamic--elastic coupling arising from volume conservation.
  Our theory reproduces the observed dependence of phase transition temperature and domain size on elastomer stiffness in isotropically swollen elastomers.
  In the companion paper, we investigate the effects of anisotropic swelling and inhomogeneous elastic moduli.
\end{abstract}

\maketitle

\section{Introduction}
\label{sec:introduction}

Elasticity and phase separation are ubiquitously interlinked in many condensed matter systems.
Examples include polymer blends and gels, where elastic networks within the material constrain compositional fluctuations~\cite{gennes1979,panyukov1996,peleg2007,tateno2021}; biological cells, where elasticity is known to regulate the formation of biomolecular condensates~\cite{hyman2014,tanaka2022,zwicker2025}; and metallic alloys, where elastic fields mediate long-range interactions between phase-separated domains~\cite{onuki1989,onuki2002}.
In these systems, elasticity can modify domain morphologies~\cite{loudet2000,nishimori1990}, alter phase boundaries~\cite{cahn1961}, and even suppress or promote phase separation~\cite{fratzl1999}.
Understanding how elasticity affects phase separation is, therefore, essential for controlling the structure and behavior of a wide variety of physical systems.

A recent experiment~\cite{fernandez-rico2024} demonstrated how the interplay between elasticity and phase separation can be harnessed to create patterned elastomers with complex morphologies.
In these experiments, crosslinked polydimethylsiloxane (PDMS) elastomers are first swollen with a solvent and then cooled to induce phase separation.
Rather than undergoing macroscopic demixing into solvent-rich and polymer-rich phases, the elastomer develops stable microphases with characteristic sizes of a few {\textmu}m, a phenomenon termed {\em elastic microphase separation} (EMPS).
For elastomers of Young's modulus $Y$, the characteristic domain size scales as $Y^{-1/2}$, and the phase transition temperature decreases linearly with $Y$.
Based on these observations, it was proposed that EMPS arises from the distinct length scales over which elastic and thermodynamic interactions operate~\cite{fernandez-rico2024}.
EMPS bears some resemblance to earlier observations of phase separation and critical density fluctuations in certain gels when cooled~\cite{tanaka1977,tanaka1978,li1989,onuki1993}.
However, unlike the elastomers studied in Ref.~\cite{fernandez-rico2024}, those gels do not exhibit microphase separation and instead undergo spinodal decomposition.

The first theoretical attempt~\cite{fernandez-rico2024} to explain EMPS was based on the Cahn--Larch\'{e} free energy~\cite{fratzl1999,onuki1989a} for metallic alloys, but it could not account for microphase formation.
Subsequent work introduced a phenomenological Coulomb-like interaction term, which produced microphase-separated domains in numerical simulations~\cite{oudich2026}.
Alternative approaches based on linear and nonlinear nonlocal elasticity have also been proposed, including a one-dimensional numerical model that captures the experimental scaling trends for deep temperature quenches and high stiffnesses~\cite{qiang2024,paulin2026}.
Broader questions concerning solvent--network equilibrium have also been explored, and strain stiffening has been shown to enable phase coexistence between the polymer network and the solvent~\cite{fernandez-rico2026,wang2025}.
Even more recently, motivated by EMPS, more general studies on nonlocal pattern formation~\cite{thewes2026} and on poroelastic coarsening and arrest in gels~\cite{safran2026} have also appeared.

In our earlier contribution towards EMPS modeling~\cite{mannattil2025}, we developed a three-dimensional (3D) phase-field model and analytically derived the domain-size scaling as observed in experiments.
Our framework is grounded in theoretical approaches to explain density fluctuations and spinodal composition in gels~\cite {onuki1993,panyukov1996,onuki1999,onuki2002} that were subsequently confirmed experimentally.
Furthermore, the transition temperature and phase behavior predicted by our model showed good quantitative agreement with experimental results.

While the models and studies discussed above provide valuable insights into EMPS and, more broadly, into phase separation,
several crucial aspects of EMPS remain insufficiently understood or unexplained.
In our previous work (see also Ref.~\cite{qiang2024}), we had assumed that the elastomers undergoing EMPS obey nonlocal linear elasticity,
with a nonlocal constitutive relation between the stress and strain~\cite{eringen1977,eringen1987}.
Extending such an approach to more general settings, particularly those involving anisotropy, remains challenging and motivates the development
of a framework that captures the same physics.

In this paper, hereafter referred to as Part I, we demonstrate that the restrictive assumption of nonlocal elasticity is not required to account for EMPS.
Instead, all its key features can be explained by introducing a physically intuitive nonlocal coupling between thermodynamics and elasticity that arises from polymer volume conservation.
This places our earlier theory~\cite{mannattil2025} on a firmer foundation while enabling the analysis of anisotropic effects, which the original formulation could not address.
Anisotropic effects are the subject of the companion paper~\cite{part2}, referred to as Part II hereafter.

In Part I, where we consider only isotropically swollen elastomers, we show that our model captures the experimentally observed scaling of domain size and transition temperature with stiffness when standard results from rubber elasticity are incorporated.
We further discuss the possible morphologies that arise during microphase separation and make comments on the phase behavior in the presence of equilibrium fluctuations.
Finally, we show that the scaling behavior observed in EMPS closely parallels that predicted by de Gennes for phase separation in crosslinked polymer blends more than four decades ago~\cite{gennes1979}.

Part I is organized as follows.
Section~\ref{sec:model} introduces the theoretical framework.
General phase behavior for isotropically swollen elastomers is discussed in Sec.~\ref{sec:general},
while comparisons with experiments are presented in Sec.~\ref{sec:experiments}.
Section~\ref{sec:degennes} compares EMPS with phase separation in crosslinked polymer blends,
and Sec.~\ref{sec:conclusion} summarizes the main findings.

\begin{figure}
  \centering\includegraphics{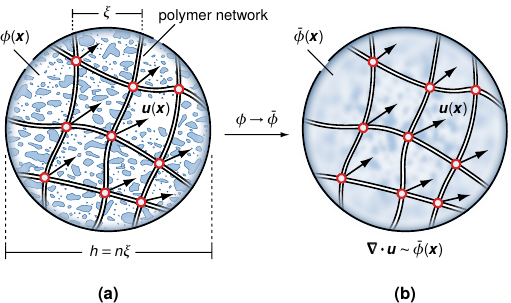}
  \caption{(a) The volume fraction of a discrete polymer network, represented as a continuum field $\phi(\bm{x})$,
    exhibits a pronounced coarseness at intermolecular length scales.
    However, only those variations in $\phi$ that persist beyond a characteristic length scale $h$ produce appreciable network deformations.
    We describe the deformations using a mesoscopic displacement field $\bm{u}(\bm{x})$.
    (b) To capture this separation of scales, $\phi$ is replaced
    with a further coarse-grained version $\bar{\phi}$ in the material-conservation relation,
    Eq.~\eqref{eq:matcons}, that links $\bm{u}$ and $\phi$.
    This associates $\bm{u}$ only with variations in $\phi$ on length scales larger than $h$.
    For isotropically swollen elastomers, we take $h = n\xi$, where $\xi$ is the end-to-end distance between adjacent crosslinks in the polymer network and $n$ is the average number of crosslinks we coarse-grain over.
  }
  \label{fig:blur}
\end{figure}

\section{Model}
\label{sec:model}

We consider a charge-neutral elastomer consisting of a crosslinked polymer network isotropically swollen by a solvent.
Because swelling results from coupled thermodynamic and elastic effects and can involve large deformations, swollen elastomers are generally described using nonlinear elasticity~\cite{flory1953,treloar1975}.
Here, we do not model the swelling process itself and we assume that the system has already reached thermodynamic equilibrium.
Furthermore, by studying small fluctuations about an already swollen reference state, we can linearize the underlying nonlinear theory and use linear elasticity.
Within this linear description, at a given temperature $T$, the system is characterized by two continuum fields shown in Fig.~\ref{fig:blur}(a): the volume fraction $\phi$ and the displacement field $\bm{u}$ of the polymer network~\cite{doi2009}.
Immediately after swelling, the network volume fraction $\phiz$ is a spatially uniform constant.
Then, upon lowering the temperature, the local volume fraction $\phi(\bm{x})$ deviates from $\phiz$.

We assume that the polymer--solvent system in the swollen elastomer is close to a critical point $(\phic, \Tc)$, where $\phic$ is the critical volume fraction and $\Tc$ the critical temperature.
After defining the order parameter $\psi(\bm{x}) = \phi(\bm{x}) - \phic$, we express the total free energy $\mathscr{F}$ as
\begin{equation}
  \mathscr{F}[\psi, \bm{u}] = \mathscr{F}_{\text{GL}}[\psi] + \mathscr{F}_{\text{el}}[\bm{u}],
  \label{eq:free_total}
\end{equation}
where $\mathscr{F}_{\text{GL}}$ is a Ginzburg--Landau free-energy accounting for network-solvent phase separation.
The second term, $\mathscr{F}_{\text{el}}[\bm{u}]$, is the elastic energy associated with the network deformation,
to be discussed in more detail momentarily.

The free energy $\mathscr{F}_{\text{GL}}$, has the form
\begin{equation}
  \mathscr{F}_{\text{GL}}[\psi] = \int \dd^{3}x\,\left[f(\psi) + \frac{1}{2}\kappa\abs{\grad\psi}^{2} + \eta(\psi - \psiz)\right].
  \label{eq:free_GL}
\end{equation}
The coordinate $\bm{x} = (x, y, z)$ in 3D space is an Eulerian coordinate describing the current state of the elastomer.
In addition, the bulk free-energy density $f(\psi)$ is written as an expansion about the critical point to quartic order,
retaining only even powers in $\psi$, i.e.,
\begin{equation}
  f(\psi) = \frac{1}{2}a(T-\Tc)\psi^{2} + \frac{1}{4}b\psi^{4},
  \label{eq:landau}
\end{equation}
where $a$ and $b$ are positive phenomenological constants.
Bulk energy densities of this form are commonly employed to describe swelling and deswelling of polymer networks~\cite{onuki1999,yamaue2000,yamaue2004}.
All thermodynamic contributions from polymer-solvent interactions are implicitly included in $f(\psi)$.
An advantage of the Landau free energy $f$ is that, for $b>0$, it has a stable double-well form.
The quartic term $\psi^{4}$ ensures thermodynamic stability and avoids limitations of mean-field free energies that lack an explicit double-well structure~\cite{fernandez-rico2026}.

The above free energy $\mathscr{F}_{\text{GL}}$, Eq.~\eqref{eq:free_GL}, also contains a squared-gradient contribution involving an interfacial parameter
$\kappa > 0$, which penalizes spatial inhomogeneities in $\psi$.
Finally, the chemical potential $\eta$ in $\mathscr{F}_{\text{GL}}$ acts as a Lagrange multiplier and constrains the spatial average of $\psi$
to a constant mean $\psiz = \phiz - \phic$, thereby conserving the total volume of the system.

As the swollen polymer is assumed to follow conventional linear elasticity, its elastic energy is
\begin{equation}
  \mathscr{F}_{\text{el}}[\bm{u}] = \frac{1}{2}\int\dd^{3}x\,\left[\lambda(\trace\strain)^{2} + 2\mu\trace(\strain^{2})\right],
  \label{eq:free_el}
\end{equation}
where $\strain = \frac{1}{2}[\grad\bm{u} + (\grad\bm{u})\trans]$ is the linearized strain, with $(\grad \bm{u})\trans$ being the transpose of $\grad \bm{u}$.
The Lam\'{e} moduli $\lambda$ and $\mu$ depend on the polymer volume fraction $\phiz$ of the elastomer in the swollen state.
The dependence is typically weak, and since the elastomer is close to criticality, we take $\phiz \approx \phic$ in the estimates below.
For an isotropically swollen elastomer composed of Gaussian chains, we have~\cite{onuki1993}
\begin{equation}
  \lambda = \nu\kbt\left(\phic - \phic^{1/3}\right),
  \quad
  \mu = \nu\kbt\phic^{1/3}.
\end{equation}
Above, $k_{\text{B}}$ is the Boltzmann constant and $\nu$ denotes the density of network strands,
namely the number of polymer segments connecting neighboring crosslinks per unit volume.
A strand here refers to the portion of a chain between two successive crosslinks.

Finally, it is also worth noting that linear theories of gels~\cite{doi2009,onuki1993,dimitriyev2019,jia2021}
are sometimes formulated entirely in terms of the displacement field $\bm{u}$.
In these theories, thermodynamic contributions that are quadratic in the strain are absorbed
into the first Lam\'{e} modulus $\lambda$ by adding an osmotic term $\phiz^{2}f''(0)$.
By contrast, our framework treats thermodynamic and elastic effects as operating at distinct length scales.
Consequently, we do not use osmotic elastic moduli in our analysis.

\subsection*{Nonlocal effects via volume conservation}

The total free energy in Eq.~\eqref{eq:free_total} is written as a sum of thermodynamic
and elastic effects, which depend on the fields $\psi$ and $\bm{u}$, respectively.
Although both $\psi$ and $\bm{u}$ describe the same underlying discrete polymer network,
they are defined at different length scales and serve distinct physical roles.
Specifically, interface formation and thermodynamic interactions, described by $\psi$,
occur at intermolecular length scales of a few \r{A},
while the elastic energy is dominated by those deformations $\bm{u}$ occurring above
a characteristic mesoscopic ({\textmu}m) scale (see Fig.~\ref{fig:blur}).

The fields $\psi$ and $\bm{u}$ are not mutually independent.
They are linked by the requirement that the total volume of the polymer network be conserved during phase separation.
Recalling that $\bm{x}$ is an Eulerian coordinate describing the current configuration of the elastomer,
conservation of polymer volume over an arbitrary subvolume yields
\begin{equation}
  \int \dd^{3}x\,\left[1 - \div\bm{u}\right]\phiz =
  \int \dd^{3}x\, \bar{\phi}(\bm{x}),
  \label{eq:volcons}
\end{equation}
where the Jacobian correction $1 - \div\bm{u}$ accounts for the linear-order change in the local volume measure due to the deformation.
Although the above equality is well known in continuum theories of gels~\cite{onuki1988,onuki1993,onuki2002,qiang2024,mannattil2025}, in Eq.~\eqref{eq:volcons} the second integral is written in terms of $\bar{\phi}(\bm{x})$ rather than $\phi(\bm{x})$.
The field $\bar{\phi}(\bm{x})$ is a coarse-grained polymer volume fraction with its short-wavelength fluctuations suppressed.
This ensures that the displacement field $\bm{u}$ is coupled only to those variations in $\phi$ occurring on comparable length scales.

There are several possible ways to choose $\bar{\phi}(\bm{x})$.
For convenience, we obtain it by filtering $\phi(\bm{x})$ using an isotropic Gaussian kernel $K(\bm{x}) = (2\pi h^{2})^{-3/2}\exp(-\frac{1}{2}\abs{\bm{x}}^{2}/h^{2})$, thereby preferentially retaining its long-wavelength components:
\begin{equation}
  \bar{\phi}(\bm{x}) =   \int\dd^{3}{x'}\, K(\bm{x} - \bm{x}')\,\phi(\bm{x}').
  \label{eq:blur}
\end{equation}
In the above equation, the length scale $h$ controls the extent to which $\bar{\phi}$ is coarse-grained again.
It is related to the polymer network's mesh size, as will be discussed in more detail in Sec.~\ref{sec:experiments} (see also Fig.~\ref{fig:blur}).
Our earlier work~\cite{mannattil2025} also used a similar Gaussian kernel to impose a nonlocal stress--strain relation.
However, the physical basis of such a constitutive relation is not transparent and limits its generalizability.
For that reason, here we retain the standard strain-energy relation, Eq.~\eqref{eq:free_el}.
Nonlocality instead arises naturally from volume conservation, which couples the fields $\bm{u}$ and $\psi$ defined at distinct length scales.

As Eq.~\eqref{eq:volcons} holds for arbitrary domains, we can equate its two integrands. Then, upon
further expanding around the critical volume fraction $\phi = \phic$, we obtain a nonlocal material conservation relationship of the form
\begin{equation}
  \div\bm{u} = -\phic^{-1}\bar{\psi}(\bm{x}) + \mathcal{O}(\psiz) + \mathcal{O}(\psi^{2}),
  \quad
  \label{eq:matcons}
\end{equation}
where $\bar{\psi}=\bar{\phi}-\phic$.

For the remainder of the analysis, it is convenient to work in Fourier space.
Fourier transforming Eqs.~\eqref{eq:blur} and \eqref{eq:matcons} we obtain (to lowest order)
\begin{equation}
  i\bm{q}\cdot\bm{u}_{\bm{q}} = -\phic^{-1}\psi_{\bm{q}}\,
  \mathrm{e}^{-\frac{1}{2}h^{2}q^{2}},
  \label{eq:matcons_fourier}
\end{equation}
where $\bm{u}_{\bm{q}} = \int \dd^{3}x\, \mathrm{e}^{-i\bm{q}\cdot\bm{x}}\, \bm{u}(\bm{x})$ and $\psi_{\bm{q}}$
are the Fourier transforms of $\bm{u}(\bm{x})$ and $\psi(\bm{x})$, respectively, with $q = \abs{\bm{q}}$.
Likewise, making use of the Parseval--Plancherel theorem, the total elastic energy, Eq.~\eqref{eq:free_el}, can be written as
\begin{align}
  \mathscr{F}_{\text{el}}[\bm{u}_{\bm{q}}] & = \frac{1}{2}\int \frac{\dd^{3}q}{(2\pi)^{3}}\Big[(2\mu
  + \lambda)(\bm{q}\cdot\bm{u}_{\bm{q}})(\bm{q}\cdot\bm{u}_{-\bm{q}})\NN\\
                                           & \quad + \mu q^{2}\bm{u}^{\perp}_{\bm{q}}\cdot\bm{u}^{\perp}_{-\bm{q}}\Big]
                                           \label{eq:free_el_fourier}
\end{align}
with $\bm{u}_{\bm{q}}^{\perp} = \bm{u}_{\bm{q}} - q^{-2}(\bm{q}\cdot\bm{u}_{\bm{q}})\bm{q}$ being the transverse component of $\bm{u}_{\bm{q}}$.

Throughout our analysis, we assume that material transport during phase separation is diffusion dominated and does not excite elastic shear modes.
Thermodynamic--elastic coupling gives rise only to longitudinal instabilities (those along $\bm{q}$),
as thermodynamic contributions to the free energy arise exclusively through the order parameter $\psi$.
Although elastic longitudinal and shear modes can undergo mode conversion at system boundaries,
such effects are negligible when the deformations remain much smaller
than the system size---a standard assumption in the theory of gels~\cite{onuki1992}.
For the elastomers studied in Ref.~\cite{fernandez-rico2024}, for example, microphase domains
are only a few \textmu{m} across, whereas the sample size is of the order of a few cm.
Boundary-induced mode conversion is, therefore, negligible, allowing us to focus on contributions to $\mathscr{F}_{\text{el}}$
from the bulk longitudinal modes alone~\cite{onuki2002}.

Consequently, we express the total elastic energy entirely in terms of the order parameter $\psi$
using the material conservation, Eq.~\eqref{eq:matcons}.
Making use of Eq.~\eqref{eq:matcons_fourier} and discarding the transverse shear modes, we find  that the total elastic energy is
\begin{equation}
  \mathscr{F}_{\text{el}}[\psi] = \frac{1}{2}\int\frac{\dd^{3}q}{(2\pi)^{3}}M_{\bm{q}}\psi_{-\bm{q}}\psi_{\bm{q}},
  \label{eq:free_el_simple}
\end{equation}
where the effective ${q}$-dependent longitudinal modulus is
\begin{equation}
  M_{\bm{q}} = \frac{\nu\kbt}{\phic^{2}}\left(\phic + \phic^{1/3}\right)\mathrm{e}^{-h^{2}q^{2}}.
  \label{eq:M_isotropic}
\end{equation}
Note that $M_{\bm{q}}$ has a $q$-dependence owing solely to the nonlocal nature of the material conservation relation, Eq.~\eqref{eq:matcons}.

\section{General phase behavior}
\label{sec:general}

\subsection{Linear stability analysis}

For linear stability analysis, we set $\psi = \psiz + \delta\psi$ in the total free energy
$\mathscr{F} = \mathscr{F}_{\text{GL}} + \mathscr{F}_{\text{el}}$, with $\mathscr{F}_{\text{GL}}$ and $\mathscr{F}_{\text{el}}$
given by Eqs.~\eqref{eq:free_GL} and \eqref{eq:free_el_simple}, respectively.
After omitting the nonquadratic terms in the free energy, which are not relevant for linear stability analysis, the total Gaussian (quadratic)
free energy of the modulations $\delta\psi$ takes the form
\begin{equation}
  \mathscr{F}_{\text{G}}[\delta\psi] = \frac{1}{2}\int\frac{\dd^{3}q}{(2\pi)^{3}}\,F_{\bm{q}}\delta\psi_{\bm{q}}\delta\psi_{-\bm{q}},
  \label{eq:free_fourier}
\end{equation}
with $F_{\bm{q}}$ being the Fourier transform of the effective binary interaction in $\delta \psi$, given by
\begin{equation}
  F_{\bm{q}} = a(T-\Tc) + 3b\psiz^{2} + \kappa q^{2} + M_{\bm{q}},
  \label{eq:free_binary}
\end{equation}
and the longitudinal modulus $M_{\bm{q}}$ is given by Eq.~\eqref{eq:M_isotropic}.
Because $M_{\bm{q}} \sim \mathrm{e}^{-h^{2}q^{2}}$, the elastic energy cost
can be reduced by preferentially enhancing modulations at short wavelengths (large $q$).
By contrast, the interfacial contribution $\kappa q^{2}$ penalizes rapid spatial variations and, therefore, favors long-wavelength
(small $q$) modulations. The competition between these two opposing tendencies can, in principle,
lead to the stabilization of a spatially modulated phase at an intermediate wavenumber~\cite{gennes1979,andelman1987,souza2026}.

During a temperature quench, the onset of this instability occurs at the temperature where $F_{\bm{q}}$ first vanishes.
If $F_{\bm{q}} \lesssim 0$ for some $q$, the corresponding $q$-modes become energetically favorable and grow,
thereby continuously lowering the free energy until the quartic term in $\mathscr{F}_{\text{GL}}$ stabilizes their growth.
Additionally, the most unstable wavenumber $\qm$ is the wavenumber
at which $F_{\bm{q}}$ has a minimum, since it is this mode that first becomes
unstable during the quench~\cite{gennes1979,leibler1987,seul1995}.
This wavenumber sets both the emergent length scale and the form of the resulting modulation.
Consequently, instead of a macroscopic phase separation into solvent-rich and polymer-rich domains,
we expect the emergence of a periodic pattern of alternating solvent-rich and polymer-rich regions.

Minimizing $F_{\bm{q}}$ in Eq.~\eqref{eq:free_binary}
with respect to $q$, we see that the most unstable wavenumber $\qm$ is
\begin{equation}
  \qm^{2} = h^{-2}\ln\gammaz.
  \label{eq:qm_0}
\end{equation}
where the dimensionless parameter $\gammaz$ is defined as
\begin{equation}
  \gammaz = \frac{M_{0}h^{2}}{\kappa}
  \label{eq:elastocapillary}
\end{equation}
and $M_{0}=M_{\bm{q}\to 0}$ from Eq.~\eqref{eq:M_isotropic} is the long-wavelength modulus
of an isotropically swollen elastomer, given by
\begin{equation}
  M_{0} = \frac{\nu\kbt}{\phic^{2}}\left(\phic + \phic^{1/3}\right).
  \label{eq:M_0}
\end{equation}
Clearly, from Eq.~\eqref{eq:qm_0}, we see that a real $\qm$ exists only if $\gammaz > 1$.

The dimensionless parameter $\gammaz$ is analogous to an elastocapillary number and measures the relative importance of elastic and interfacial contributions to the free energy.%
\footnote{Our definition of the elastocapillary number $\gammaz$ differs from the more conventional
  one~\cite{ronceray2022,liu2019}, which is defined as the ratio of capillary (interfacial) effects to elastic effects.}
When $\gammaz < 1$, interfacial costs dominate, and the system does not undergo microphase separation.
Instead, a bulk linear instability develops below the temperature $\Tc - a^{-1}[3b\psiz^{2} + M_{0}]$ and the system undergoes a conventional macrophase separation into polymer-rich and solvent-rich phases.

When $\gammaz > 1$, elastic costs dominate, favoring the formation of many stable, finite-size domains, resulting in microphase separation.
In this case, setting $q = \qm$ in $F_{\bm{q}} = 0$, we find the temperature $\Tm$ at which microphase separation occurs to be
\begin{equation}
  \Tm(\psiz) = \Tc - a^{-1}\left[3b\psiz^{2} + M_{0}\gammaz^{-1}(1 + \ln\gammaz)\right].
  \label{eq:Tm_iso}
\end{equation}
From the above equation, we see that elastic effects---quantified by the longitudinal modulus $M$---lowers the microphase separation temperature $\Tm$.
Such a downward shift of the transition temperature due to elasticity is well documented (for example, in metallic alloys~\cite{cahn1961}).
It shows that elasticity increases the temperature range over which the system remains stable.

The onset of microphase separation can also be gleaned from the static structure factor $S(\bm{q}) = \langle\delta\psi_{\bm{q}}\,\delta\psi_{-\bm{q}}\rangle$.
At temperatures above $\Tm$, the scattering intensity arising from equilibrium fluctuations $\delta\psi$ is proportional to $S(\bm{q})$.
Within the Gaussian approximation, $S(\bm{q})$ can be evaluated (up to multiplicative constants) as~\cite{onuki2002}
\begin{align}
  S(\bm{q}) &= \frac{\int \mathcal{D}[\delta \psi_{\bm{q}}]\,\mathcal{D}[\delta \psi_{-\bm{q}}]\,
  \delta\psi_{\bm{q}}\,\delta\psi_{-\bm{q}}\,
  \mathrm{e}^{-\mathscr{F}_{\text{G}}/\kbt}}
  {\int \mathcal{D}[\delta \psi_{\bm{q}}]\,\mathcal{D}[\delta \psi_{-\bm{q}}]\,
  \mathrm{e}^{-\mathscr{F}_{\text{G}}/\kbt}}\NN\\
            &\sim F_{\bm{q}}^{-1},
  \label{eq:structure}
\end{align}
For $\gammaz > 1$, the structure factor $S(\bm{q})$ exhibits a peak at $q = \qm$, with the peak height $S(\bqm) \sim (T - \Tm)^{-1}$.
As the elastomer temperature is reduced toward $\Tm$, the scattering intensity develops a continuously growing peak at
a fixed wavenumber $\qm$, signaling the onset of instability and the formation of periodic structures.

\begin{figure}{\centering\includegraphics{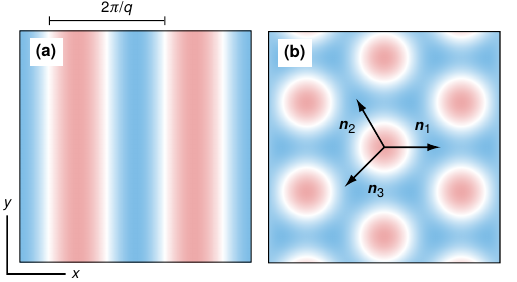}}
  \caption{Schematics of modulated phases showing (a) lamellae and (b) hexagonal phase, which also shows the unit-vector triad $\{\bm{n}_{j}\}$.
    At temperatures close to the critical temperature, the modulations are well approximated by sinusodial functions.
  In both panels, solvent rich (deficient) regions are shown in red (blue).}
  \label{fig:modulated}
\end{figure}

\subsection{Modulated phases}

In the weak-segregation regime ($T \lesssim \Tc$), where the modulations are well represented by sinusoidal oscillations, a phase diagram can be obtained analytically using the single-mode approximation~\cite{garel1982,leibler1987,andelman1987}.
Upon re-expressing the elastic free energy $\mathscr{F}_{\text{el}}$, Eq.~\eqref{eq:free_el_simple},
in real space, the total free energy $\mathscr{F} = \mathscr{F}_{\text{GL}}
  + \mathscr{F}_{\text{el}}$ for isotropically swollen elastomers becomes
\begin{equation}
  \mathscr{F} = \int\dd^{3}x\,\left[f(\psi) + \frac{1}{2}\kappa\abs{\grad\psi}^{2} + \frac{1}{2}
    M_{0}(\bar{\psi})^{2}\right].
  \label{eq:free_total_2D}
\end{equation}
Although our discussion so far has been in 3D, to keep the analysis of phase behavior tractable, we limit it to two-dimensional (2D) modulations.
In 2D, one considers two spatially modulated phases, namely, lamellar, and hexagonally symmetric phases, as well as a spatially uniform phase.
From the free-energy densities of each phase, the phase boundaries follow from a standard common-tangent construction.
To streamline the discussion below, we introduce the following dimensionless parameter
\begin{equation}
  \Sigma = b^{-1}\left[a(T-\Tc) + 3b\psiz^{2} + M_{0}\gammaz^{-1}(1 + \ln\gammaz)\right].
  \label{eq:gamma_param}
\end{equation}

\paragraph{Uniform phase} For the spatially uniform phase with $\psi(\bm{x}) = \psiz$, the free-energy density takes the form
\begin{equation}
  f_{\text{U}}(\psiz, T) = \frac{1}{2}\left[a(T-\Tc) + M_{0}\right]\psiz^{2} + \frac{1}{4}b\psiz^{4}.
  \label{eq:free_unif_iso}
\end{equation}

\paragraph{Lamellar phase} To represent the morphology of the lamellar phase, we assume that the order parameter
\begin{equation}
\psi(\bm{x}) = \psiz + \delta\psi= \psiz+ A\cos(qx),
\end{equation}
representing a unidirectional modulation (chosen arbitrarily along the $x$-direction) with amplitude $A$ and wavenumber $q$.
See Fig.~\ref{fig:modulated}(a) for an illustration.
Inserting this expression into the total free energy, Eq.~\eqref{eq:free_total_2D}, and minimizing it with respect to both $q$ and $A$ yields
the free-energy density $f_{\text{L}}$ of the lamellar phase as
\begin{equation}
  f_{\text{L}}(\psiz, T) = f_{\text{U}}(\psiz, T) - \frac{b}{6}\Sigma^{2},
  \label{eq:free_lam_iso}
\end{equation}
with the wavenumber $\qm$ and amplitude $A_{\text{m}}$ of the modulation being
\begin{equation}
  \qm^{2} = h^{-2}\ln\gammaz,
  \quad
  \Am^{2} = -\frac{4}{3}\Sigma.
  \label{eq:qmL}
\end{equation}
Clearly, the above $\qm$ is the same as the one derived in Eq.~\eqref{eq:qm_0} using linear stability analysis.
As the modulation amplitude $\Am$ must be real for the lamellar phase to exist, the condition $\Sigma=0$ in Eq.~\eqref{eq:gamma_param}
provides an estimate for the onset temperature of microphase separation, and we recover the expression in Eq.~\eqref{eq:Tm_iso}.

\paragraph{Hexagonal phase} For the hexagonal phase, we consider modulations in the $(x,y)$ plane of the form~\cite{andelman1987,elder2004}
\begin{equation}
  \delta\psi(\bm{x}) = A\sum_{j=1}^{3}\cos\left(q\,\bm{n}_{j}\cdot\bm{x}\right),
\end{equation}
with the 2D unit vectors $\bm{n}_{j}$ satisfying $\sum_{j=1}^{3} \bm{n}_{j} = 0$.
The hexagonal phase is illustrated in Fig.~\ref{fig:modulated}(b).
Following the same procedure as for the lamellar phase, the solution $\psiz + \delta\psi(\bm{x})$ is substituted into Eq.~\eqref{eq:free_total_2D},
which, after minimization, yields the free-energy density of the hexagonal phase as
\begin{equation}
  f_{\text{H}}(\psiz, T) = f_{\text{U}}(\psiz, T) - \frac{3b}{64}A_{\text{m},\pm}^{2}\left(\psiz A_{\text{m},\pm} - 2\Sigma\right),
\end{equation}
with $\qm$ as in Eqs.~\eqref{eq:qm_0} and \eqref{eq:qmL}, and $A_{\text{m}}$ given by
\begin{equation}
  A_{\text{m},\pm} = \frac{4}{5}\left[\psiz \pm \left(\psiz^{2} - \frac{5}{3}\Sigma\right)^{1/2}\right].
\end{equation}
Modulations with amplitude $A_{\text{m},+}$ correspond to the conventional hexagonal phase, while $A_{\text{m},-}$ describes the inverted hexagonal phase, with polymer-rich, hexagonally symmetric domains dispersed in a solvent-rich matrix.

Phase coexistence curves are determined using the common-tangent construction, which enforces equality of both the chemical potential and the osmotic pressure between the different phases.
This leads to the coupled system of equations
\begin{align}
  \frac{\partial f_{i}}{\partial \psi_{0,i}} & = \frac{\partial f_{j}}{\partial \psi_{0,j}}, \\
  \psi_{0,i}\left(\frac{\partial f_{i}}{\partial \psi_{0,i}}\right) - f_{i}
                                             & =
  \psi_{0,j}\left(\frac{\partial f_{j}}{\partial \psi_{0,j}}\right) - f_{j}.
\end{align}
The indices $i$ and $j$ label any coexisting pair among the uniform, lamellar, hexagonal, and inverted hexagonal phases.
By numerically solving these equations, one obtains the phase diagram as a function of temperature $T$ and mean polymer volume fraction $\phiz$.

\begin{figure}
  \centering\includegraphics{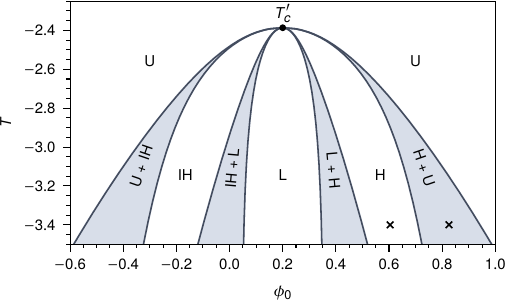}
  \caption{Phase diagram in the polymer volume fraction--temperature plane $(\phiz, T)$,
    for an isotropically swollen elastomer with elastocapillary number [see Eq.~\eqref{eq:elastocapillary}], $\gammaz = 4$.
    For illustrative purposes, we set the critical parameters to $(\phic, \Tc) = (0.2, 0)$.
    All other free-energy parameters are set to unity.
    Elastic effects shift the critical temperature to $\Tc' = \Tm(0) = -(1+\ln 4) \approx -2.386$ [Eq.~\eqref{eq:Tm_iso}] below which three phases appear: lamellar (L), hexagonal (H), and inverted hexagonal (IH).
    Only the uniform (U) phase appears above $\Tc'$.
    The phase coexistence regions separated by binodals (solid curves) are shaded in grey.
    The two crosses in the lower right corner indicate the $(\phiz, T)$ values used later in Fig.~2 of Part II~\cite{part2}.
    The part of the phase diagram with $\phiz < 0$ is shown for illustration purposes alone and does not constitute a physical scenario for the problem treated here.}
  \label{fig:isotropic}
\end{figure}

A representative phase diagram constructed from the above free-energy expressions for 2D modulations is shown in Fig.~\ref{fig:isotropic} in the $(\phiz,T)$ plane.
Here, we have set the long-wavelength modulus to $M_{0} = 4$ and all other free-energy parameters to unity, yielding an elastocapillary number of $\gammaz = 4$.
In polymer networks, we expect $h$ to be micron-scale or less, while interfaces form over a few \r{A}, so $\mathcal{O}(\gammaz)\sim \text{1--10}$
corresponds to very soft networks with stiffness of a few kPa, comparable to certain hydrogels~\cite{luo2022}.

In the vicinity of the critical point, three principal phases appear: a uniform phase, a hexagonal or droplet phase consisting of solvent-rich domains embedded in a polymer-rich matrix, and a lamellar phase made up of alternating solvent-rich and polymer-rich bands.
An additional inverted-hexagonal phase is also present, where solvent-poor regions form isolated domains within a solvent-rich background.
In between these phases, four regions of phase coexistence appear.

Except at the critical point---where a second-order phase transition from the uniform to the lamellar phase can occur---all phase transitions displayed in Fig.~\ref{fig:isotropic} are first order.
The topology of the phase diagram closely resembles that of other systems with modulated phases such as ferromagnets~\cite{garel1982}, block copolymers~\cite{fredrickson1987}, membranes~\cite{leibler1987,yu2025a}, monolayers~\cite{andelman1987}, etc., which is not surprising because their long-wavelength behavior is similar, as we saw from Eq.~\eqref{eq:free_binary_simple}.

\subsection{Lifshitz behavior}

For small $q$ and $\gammaz \gtrsim 1$, we can expand $F_{\bm{q}}$ in Eq.~\eqref{eq:free_binary} to $\mathcal{O}(q^{4})$, resulting in
\begin{equation}
  F_{\bm{q}} = a(T-\Tc) + M_{0} + 3b\psiz^{2} + (1 - \gammaz)\kappa q^{2} + \frac{1}{2}\gammaz\kappa h^{2}q^{4}.
  \label{eq:free_binary_simple}
\end{equation}
The form of $F_{\bm{q}}$ considered here commonly appears in a wide variety of pattern-forming systems.
Examples include those described by Landau--Brazovskii or Swift--Hohenberg-type free energies, such as bilayer membranes~\cite{leibler1987}, block copolymers~\cite{fredrickson1987}, Langmuir monolayers~\cite{leibler1987}, phase-field crystals~\cite{elder2002}, etc.
In the presence of fluctuations, these systems can display the so-called Lifshitz behavior~\cite{hornreich1975,chaikin1995}.

To understand the effect of equilibrium fluctuations, we consider the structure factor
$S(\bm{q}) \sim F_{\bm{q}}^{-1}$, Eq.~\eqref{eq:structure}, which is of the form%
\begin{equation}
  S(\bm{q}) = \frac{S(0)\tau}{\tau + 2(1 - \gammaz)q^{2} + \gammaz h^{2}q^{4}},
  \label{eq:sfactor_approx}
\end{equation}
where $\tau$ is an effective temperature given by
\begin{equation}
\label{eq:tau}
  \tau = 2\kappa^{-1}[a(T-\Tc) + 3b\psiz^{2} + M].
\end{equation}
The inverse Fourier transform of $S(\bm{q})$ is the real-space correlation function,
\begin{equation}
  G(\bm{x}) = \langle\delta\psi(\bm{x})\delta\psi(0)\rangle,
  \label{eq:g}
\end{equation}
which quantifies the spatial correlation of composition fluctuations.

\begin{figure}
  \centering\includegraphics{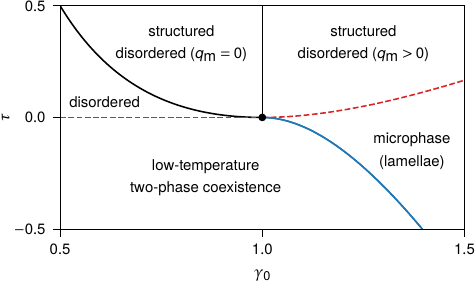}
  \caption{Phase diagram in the $(\gammaz,\tau)$ plane for $h=1$,
    where $\tau$ is the effective temperature defined in Eq.~\eqref{eq:tau},
    and $\gammaz$ is the elastocapillary number, Eq.~\eqref{eq:elastocapillary}.
    Second-order phase transition boundaries are shown as dashed red curves and correspond to $\tau = (1 - \gammaz)^{2}/\gammaz$ for $\gammaz>1$, and to $\tau=0$ for $\gammaz<1$.
    The solid blue curve, given by $\tau=-(2+\sqrt{6})(1-\gammaz)^{2}/\gammaz$, marks a first-order triple line separating the lamellar phase from the two-phase coexistence region.
    The solid black curve defined by $\tau = (1 - \gammaz)^{2}/\gammaz$ for $\gammaz < 1$ is the disorder line, which distinguishes the structured-disorderd regime with $\qm \neq 0$ fluctuations from the conventional disordered phase.
    The vertical black line at $\gammaz=1$ is the Lifshitz line, terminating at the Lifshitz point (black dot).
    Both the disorder and Lifshitz lines represent regime crossovers rather than thermodynamic phase transitions.}
  \label{fig:lifshitz}
\end{figure}

To streamline the discussions below, we introduce the dimensionless parameter
\begin{equation}
  \chi = \frac{1 - \gammaz}{\sqrt{\gammaz \tau h^2}}.
  \label{eq:chi}
\end{equation}
The correlation function associated with Eq.~\eqref{eq:sfactor_approx} takes the following form when $\abs{\chi} < 1$~\cite{teubner1987,komura2007}:
\begin{equation}
  G(\bm{x}) = G(0)\,\mathrm{e}^{-\abs{\bm{x}}/\zeta}\sinc\left(\frac{2\pi \abs{\bm{x}}}{d}\right).
  \label{eq:correlation}
\end{equation}
The above correlation function describes damped oscillations with correlation length $\zeta$ and period $d$, given by
\begin{equation}
  \zeta^{4} = \frac{4\gammaz h^{2}}{\tau(1 + \chi)^{2}},\quad
  d^{4} = \frac{4\gammaz h^{2}}{\tau(1 - \chi)^{2}}.
\end{equation}
Using the above results, we can summarize the general phase behavior of elastomers in the $(\gammaz, \tau)$ plane as follows (also illustrated in Fig.~\ref{fig:lifshitz}).

\begin{table*}
\centering
\caption{Parameter definitions and their values.}
\label{tab:param}
\begin{ruledtabular}
\begin{tabular}{l l l l}
  Parameter & Description & Value\\
  \hline
  $a$ & Quadratic coefficient of the Landau free energy $f(\psi)$ [Eq.~\eqref{eq:landau}]
  & $0.025\unit\text{kPa}\unit\text{K}^{-1}$ ($25\unit\text{J}\unit\text{m}^{-3}\unit\text{K}^{-1}$)\\
  $b$ & Quartic coefficient of $f(\psi)$ & $2\unit\text{kPa}\unit\text{K}^{-1}$ ($2\times10^{3}\text{J}\unit\text{m}^{-3}$)\\
  $\Tc$ & Critical temperature of $f(\psi)$ & $70\unit{^\circ}\text{C}$ ($343\unit\text{K}$)\\
  $\phic$ & Critical volume fraction of $f(\psi)$ & 0.2\\
  $n$ & Mean number of crosslinks coarse-grained over [Eq.~\eqref{eq:cgl}] & 35 \\
  $C_{\infty}$ & Flory characteristic ratio of PDMS~\cite{rubinstein2003} & 6.8\\
  $\varrho$ & Mass density of dry PDMS~\cite{kuo1999} & $970\unit\text{kg}\unit\text{m}^{-3}$\\
  $\ell$ & Length of the PDMS monomer [--$\text{Si}(\text{CH}_{3})_{2}\text{O}$--]~\cite{mark2004} & $3.28\unit\text{\r{A}}$\\
  $\mz$ & Molecular mass of PDMS monomer & $1.23\times10^{-25}\unit\text{kg}$ ($74.2\unit\text{g}\unit\text{mol}^{-1}$)\\
  $\kbt$ & Thermal energy at $T = 300\unit\text{K}$ & $4.14\times 10^{-21}\unit\text{J}$\\
  $B$ & $C_{\infty}\varrho\ell^{2}\kbt/\mz$ [Eq.~\eqref{eq:mesh_size}] & $0.024\unit\text{kPa}\unit\text{\textmu m}^{2}$
  ($2.4\times10^{-11}\unit\text{J}\unit\text{m}^{-1}$)\\
  $\kappa$ & Interfacial parameter ($\sim \kbt/\ell$~\cite{leibler1987}) & $0.013\unit\text{kPa}\unit\text{\textmu m}^{2}$
  ($1.3\times10^{-11}\unit\text{J}\unit\text{m}^{-1}$)\\
\end{tabular}
\label{tab:parameters}
\end{ruledtabular}
\end{table*}

{\it i})~When the effective temperature $\tau < 0$ and $\gammaz < 1$, the system undergoes macrophase demixing and exists in a low-temperature coexistence phase.

{\it ii})~For $\abs{\chi} < 1$, both $\zeta$ and $d$ remain finite.
The system is then in a structure-disordered phase, characterized by fluctuating mesoscopic structures embedded within an otherwise disordered medium~\cite{komura2007}.
The oscillatory decay of $G(\bm{x})$ reflects this behavior.

{\it iii})~When $\gammaz < 1$, we have $0 < \chi < 1$, and the structure factor exhibits a maximum
only at $\qm = 0$, despite the oscillatory form of the correlation function.
As $\chi$ approaches unity, the characteristic period $d$ diverges.
The condition $\chi = 1$ defines the disorder line~\cite{komura2007}.
When $\chi > 1$, oscillations disappear and $G(\bm{x})$ decays monotonically, corresponding to a fully disordered phase.
In the limit $\chi \gg 1$, the positive $q^{2}$ contribution dominates the denominator of Eq.~\eqref{eq:sfactor_approx}.
The resulting structure factor reduces to an Ornstein--Zernike form with purely exponential decay.

{\it iv})~The condition $\chi = 0$ (equivalently, $\gammaz = 1$) marks the Lifshitz line~\cite{hornreich1975}, which terminates at the Lifshitz point $(\gammaz, \tau) = (1, 0)$.
For $\gammaz > 1$ and $-1 < \chi < 0$, the structure factor develops a peak at $\qm^{2} = h^{-2}(1 - \gammaz^{-1})$,
signaling the presence of fluctuating microstructures.
As $\chi \to -1$, the correlation length diverges, indicating the emergence of long-range order.
Finally, when $\chi < -1$ stable microphases appear.

Because Eq.~\eqref{eq:free_binary_simple} belongs to the Swift--Hohenberg class of free energies,
critical fluctuations (beyond the mean-field treatment presented here)
are expected to modify the phase behavior near the order--disorder phase transition ~\cite{brazovskii1975}.
In particular, the second-order critical point is replaced by a line of fluctuation-induced first-order phase transitions,
an effect known to be important in block copolymers~\cite{koga1999,komura2008}.
While the mean-field phase diagrams we have presented in Figs.~\ref{fig:isotropic} and \ref{fig:lifshitz} should describe
the phase behavior away from the critical point, fluctuations may result in direct phase transitions between
disordered and ordered phases, including the gyroid phase in 3D systems~\cite{hamley1997}.

\section{Comparison with experiments}
\label{sec:experiments}

The results reported in Ref.~\cite{fernandez-rico2024} are primarily for isotropically swollen elastomers with uniform stiffnesses,
and we only consider such systems in the present work (Part I).

\subsection{Rubber elasticity}

The microphase domain size and the phase transition temperature in Ref.~\cite{fernandez-rico2024}
are reported in terms of the Young's modulus $Y$ of the (dry) PDMS elastomer before swelling.
The Young's modulus of the dry elastomer is related to the strand density $\nu$ via~\cite{tanaka2011}
\begin{equation}
  Y = 3\nu\kbt.
  \label{eq:youngs_dry}
\end{equation}

In elastomers and gels, a common measure of the network length scale is the end-to-end distance
of the strands between crosslinks~\cite{parrish2017,richbourg2020}.
If the strands are modeled as freely jointed chains with a Flory characteristic ratio $C_{\infty}$,
the root-mean-square end-to-end distance $\xi$ in the unswollen state follows from
the standard relation~\cite{rubinstein2003,tanaka2011,lodge2020}
\begin{equation}
  \xi^{2} = C_{\infty}N\ell^{2}.
  \label{eq:mesh_size_basic}
\end{equation}
In this expression, $N$ denotes the degree of polymerization, corresponding to
the number of monomers along a strand, while $\ell$ is the length
of a single PDMS monomer [--$\text{Si}(\text{CH}_{3})_{2}\text{O}$--].
For PDMS and other polymers with siloxane backbones, one can take $\ell$ to be equal
to twice the Si--O bond length of $1.64\unit\text{\r{A}}$~\cite{mark2004}.

In polymer networks with strands of mean molecular mass $\ms$ and monomers of mass $\mz$, the mean degree of polymerization of the strand between crosslinks is $N = \ms/\mz$~\cite{lodge2020}.
The strand mass $\ms$ can be determined from $Y$ in Eq.~\eqref{eq:youngs_dry} by noting that the mean strand density $\nu = \varrho/\ms$, where $\varrho$ is the mean mass density of the dry elastomer.
This yields an estimate for $N$ in terms of the Young's modulus $Y$ as
\begin{equation}
  N = \frac{\ms}{\mz} = \frac{3\varrho\kbt}{Y\mz}.
  \label{eq:repeat_units}
\end{equation}
For the PDMS elastomers considered in Ref.~\cite{fernandez-rico2024}, we find that $N$ varies from around $N = 10^{2}$ (for $Y=800\unit\text{kPa}$) to nearly $N = 10^{4}$ (for $Y=10\unit\text{kPa}$).
Using Eq.~\eqref{eq:repeat_units} in Eq.~\eqref{eq:mesh_size_basic}, the characteristic distance $\xi$
between crosslinks in an elastomer takes the form~\cite{yoo2006,parrish2017}
\begin{equation}
  \xi \approx \left(\frac{3B}{Y}\right)^{1/2}
  \enspace\text{with}\quad B = \frac{C_{\infty}\varrho\ell^{2}\kbt}{\mz}.
  \label{eq:mesh_size}
\end{equation}
The material-specific parameter $B$ carries dimensions of energy per unit length, and for PDMS, we estimate $B \approx 0.024\unit\text{kPa}\unit\text{\textmu m}^{2}$ using the values of the relevant physical parameters compiled in Table~\ref{tab:parameters}.
The mesh size predicted by Eq.~\eqref{eq:mesh_size} varies from $\xi \approx 5\unit\text{nm}$ at $Y = 800\unit\text{kPa}$ to $\xi \approx 50\unit\text{nm}$ at $Y = 10\unit\text{kPa}$.

Here, it is prudent to note that alternative estimates of the mesh size $\xi$ may exhibit scaling behavior that differs from Eq.~\eqref{eq:mesh_size}.
For example, if the network volume element is assumed to scale with the mesh size as $\xi^{3}$, the strand density can be estimated as $\nu \sim \xi^{-3}$.
Putting $\nu \sim \xi^{-3}$ in $Y = 3\nu\kbt$ leads to the scaling relation $\xi \sim (3\kbt/Y)^{1/3}$~\cite{gennes1979,haggerty1988,tsuji2018},
which is sometimes referred to as the rheological mesh size~\cite{wisniewska2018}.
However, this estimate is only valid for weakly crosslinked networks, where it is reasonable to assume
that each volume element of the network contains a single strand.
By contrast, the estimate in Eq.~\eqref{eq:mesh_size} is consistent with the central assumption of rubber elasticity---namely
that it originates from changes in the configurational entropy of strands following Gaussian statistics
with a mean-square end-to-end distance given by Eq.~\eqref{eq:mesh_size_basic}.

\begin{figure*}
  \centering\includegraphics{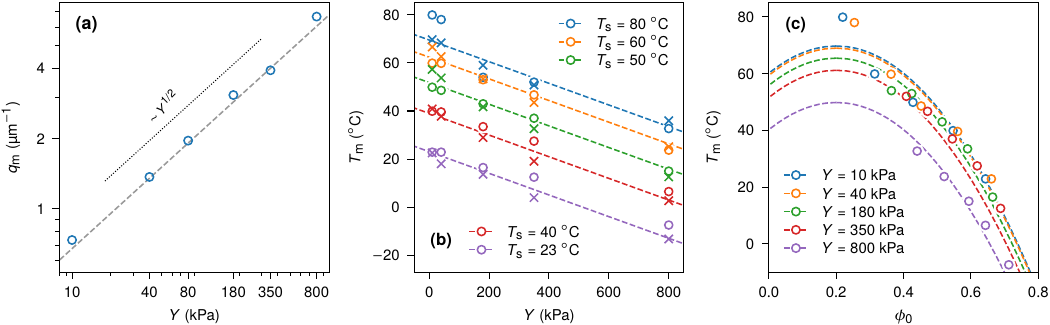}
  \caption{(a) Scattering intensity peak $\qm$ as a function of the Young's modulus $Y$ (log-log plot).
    Experimental results from Ref.~\cite{fernandez-rico2024} are shown with circles.
    The dashed line is the theoretical prediction from Eq.~\eqref{eq:qm_real}, $\qm \sim Y^{1/2}$.
    (b) Microphase separation temperature $\Tm$ as function of $Y$ for elastomers with different initial swelling temperatures
    $T_{\text{s}}$, which determine the value of the polymer volume fraction $\phiz$ at equilibrium.
    Experimental values are shown as circles, and theoretical estimates from Eq.~\eqref{eq:Tm_real} are shown as crosses.
    The dashed guidelines illustrate the linearity of $\Tm$ with $Y$.
    (c) $\Tm$ as a function of $\phiz$ for different $Y$ using Eq.~\eqref{eq:Tm_real} (dashed curves) and experimental results (circles).
    All physical parameters used in obtaining the theoretical results have been compiled in Table~\ref{tab:parameters}.
    \label{fig:scaling}
  }
\end{figure*}

\subsection{Scaling behavior and phase diagram}

From Eq.~\eqref{eq:M_0} we see that apart from the critical volume fraction $\phic$ and $\kbt$,
the long-wavelength longitudinal modulus $M_{0}$ of isotropically swollen elastomers only involves the strand density $\nu$.
One can thus use Eq.~\eqref{eq:youngs_dry} to eliminate $\nu$ and write $M_{0}$ in terms of the dry Young's modulus $Y$ as
\begin{equation}
  M_{0} = \frac{1}{3}\left(\phic^{-1} + \phic^{-5/3}\right)Y.
  \label{eq:longitudinal}
\end{equation}

The typical intermolecular spacing (roughly equal to the monomer size $\ell$) is only a few \r{A}, much smaller than the mesh size $\xi$.
As we have remarked previously, thermodynamic interactions between the polymer and the solvent occur at intermolecular length scales, whereas the polymer network can be considered as a continuous elastic medium only at length scales considerably larger than $\xi$, though proportional to it.
Based on this, for isotropically swollen elastomers, we take the coarse-graining length scale appearing in Eq.~\eqref{eq:blur} to be
\begin{equation}
  h = n \, \xi \, \phic^{-1/3} \approx n\left(\frac{3B}{Y}\right)^{1/2}\phic^{-1/3}.
  \label{eq:cgl}
\end{equation}
Here, the dimensionless factor $n=h/\xi$ can be interpreted as the average number of crosslinks included along each spatial direction during coarse-graining (see Fig.~\ref{fig:blur}).
Its precise value depends on the kernel employed in Eq.~\eqref{eq:blur} and can be determined only by comparing theoretical predictions with experimental data.
In general, broader, long-range kernels lead to smaller values of $n$.
Finally, the additional factor of $\phic^{-1/3}$ in Eq.~\eqref{eq:cgl} reflects the change in the mesh size due to isotropic swelling.

Using Eqs.~\eqref{eq:longitudinal} and~\eqref{eq:cgl}, we find the elastocapillary number $\gammaz$ of an isotropically swollen elastomer to be
\begin{equation}
  \gammaz = \frac{M_{0}h^{2}}{\kappa} = Bn^{2}\kappa^{-1}\left(\phic^{-7/3} + \phic^{-5/3}\right).
\label{eq:gammaz}
\end{equation}
As $M_{0} \sim Y$ and $h^{2} \sim \xi^{2} \sim Y^{-1}$, it is important to note that the elastocapillary number $\gammaz$
is independent of the Young's modulus $Y$.
From the parameter values compiled in Table~\ref{tab:param}, we see that $\gammaz \gg 1$, even when $n = 1$.

The wavenumber $\qm$ at which the scattering intensity peaks is obtained from Eq.~\eqref{eq:qm_0} as
\begin{equation}
  \qm^{2} = h^{-2}\ln\gammaz = Y\left(\frac{\phic^{2/3}}{3Bn^{2}}\right)\ln\gammaz.
          \label{eq:qm_real}
\end{equation}
In Fig.~\ref{fig:scaling}(a), we compare the prediction of Eq.~\eqref{eq:qm_real} with the experimental results of Ref.~\cite{fernandez-rico2024}
and find good agreement between the two.
The scaling $\qm^{2} \sim Y$ in Eq.~\eqref{eq:qm_real} is very different from the naive estimate
one might expect based on dimensional grounds alone, e.g., $\qm^{2} = (Y/\kbt)^{2/3}$.
For the experimental ranges of $Y \approx 10\text{--}800\unit\text{kPa}$, the naive estimate gives a domain size of $2\pi\qm^{-1} \approx 10\text{--}60\unit\text{nm}$, which is two orders of magnitude smaller than the experimental observations of micron-sized domains.

One can also estimate the microphase separation temperature from Eq.~\eqref{eq:Tm_iso} as
\begin{align}
  \Tm(\psiz) &= \Tc - 3ba^{-1}\psiz^{2}\NN\\
             &\quad- {Y}a^{-1}\left(\frac{\phic^{2} + \phic^{2/3}}{1 + \phic^{2/3}}\right)\left(1 + \ln\gammaz\right).
             \label{eq:Tm_real}
\end{align}
A comparison of the phase transition temperatures $\Tm$, predicted from the above equation, with the experimentally
measured ones, is showcased in Figs.~\ref{fig:scaling}(b) and~\ref{fig:scaling}(c) using all available data from Ref.~\cite{fernandez-rico2024}.
We again observe good agreement, with $\Tm$ decreasing linearly with $Y$, as predicted.

\begin{figure}
  \centering\includegraphics{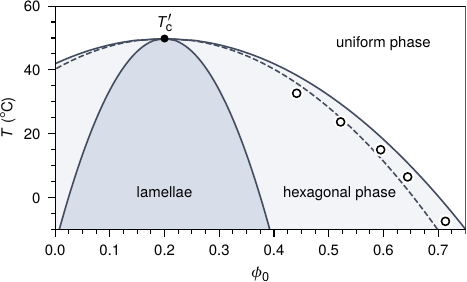}
  \caption{Phase diagram in the polymer volume fraction-temperature $(\phiz, T)$ plane for an isotropically swollen elastomer with a dry Young's modulus
    $Y = 800\,\unit{\text{kPa}}$.
    Open circles indicate experimental data taken from Ref.~\cite{fernandez-rico2024} for an elastomer with the same stiffness.
    Other material parameters are listed in Table~\ref{tab:param}.
    The binodals (phase boundaries) are shown as solid curves.
    The phase coexistence regions are extremely narrow and have not been included.
    The dashed curve corresponds to the microphase separation temperature $\Tm(\phiz)$ obtained from linear stability analysis,
    Eq.~\eqref{eq:Tm_real} [also presented in Fig.~\ref{fig:scaling}(c)].
    The binodals converge at a shifted critical point $\Tc' = \Tm(\phic)$.
  }
  \label{fig:real_isotropic}
\end{figure}

As we discussed in Sec.~\ref{sec:general}, we can construct a phase diagram in the weak-segregation limit using the single-mode approximation.
Such a phase diagram for an elastomer of stiffness $Y = 800\unit\text{kPa}$ is presented in Fig.~\ref{fig:real_isotropic}.
The phase diagram is in good agreement with experimental observations and correctly predicts the onset of microphase separation.
The first-order phase transition lines separating the various phases terminate at a critical point, $\Tc'=\Tm(\phic)$,
where a second-order transition between the uniform and lamellar phases is found.
The associated coexistence regions are extremely narrow and have therefore been omitted.
Their small widths may help explain the reported absence of hysteresis in EMPS~\cite{fernandez-rico2024}.

Experiments indicate that droplets (hexagonal phase) appear in soft elastomers with $Y \lesssim 40\unit\text{kPa}$,
whereas stiffer elastomers exhibit only channel-like structures, somewhat analogous to a lamellar phase.
Nevertheless, due to the generic topology of the theoretical phase diagram, we expect the emergence of
the hexagonal phase for off-critical temperature quenches, regardless of stiffness.
This discrepancy suggests that additional mechanisms, such as shear deformations or nonlinear effects, which have been neglected in our study,
likely govern the formation of channel-like structures in stiffer samples.

We note that, for isotropically swollen elastomers, the scaling $\qm^{2} \sim Y$ does not depend on the specific kernel used in Eq.~\eqref{eq:blur}.
For a general isotropic and normalized kernel $K(\bm{x})$, the effective longitudinal modulus,
Eq.~\eqref{eq:M_isotropic}, can be written as $M_{\bm{q}} = M_{0}K_{\bm{q}}$.
Since $K(\bm{x})$ is normalized, its Fourier transform $K_{\bm{q}}$ is dimensionless.
Furthermore, because $h$ is the only length scale entering the kernel, $K_{\bm{q}}$ can depend only on the dimensionless quantity $hq$.
Isotropy then requires $K_{\bm{q}}$ to be an even function of $\bm{q}$, which implies $M_{\bm{q}} = M_{0}K_{\bm{q}}(h^{2}q^{2})$.
Hence, the wavenumber $\qm$
at which the scattering intensity $S(\bm{q}) \sim (\kappa q^{2} + M_{\bm{q}})^{-1}$ peaks is the solution of
\begin{equation}
  1 + \gammaz K_{\bm{q}}'(h^{2}\qm^{2}) = 0.
\end{equation}
As before $M_{0} \sim Y$ and $h^{2} \sim Y^{-1}$ making the elastocapillary number $\gammaz = M_{0}h^{2}/\kappa$ independent of the Young's modulus $Y$.
Hence, any $\qm$ obtained as a solution to the above equation must scale as $\qm^{2} \sim h^{-2} \sim Y$.
A similar argument applies to the linear dependence of the phase transition temperature $\Tm$ on $Y$ as in Eq.~\eqref{eq:Tm_real}.

\section{EMPS and de Gennes' model of crosslinked blends}
\label{sec:degennes}

An early model examining the effect of elasticity on phase separation was suggested by de Gennes for crosslinked AB polymer blends~\cite{gennes1979}.
While an uncrosslinked blend undergoes macroscopic demixing below its critical temperature,
the presence of crosslinks introduces elastic effects that can stabilize microphases with a finite characteristic length scale.
De Gennes' work on blends is in the same spirit as EMPS, and despite being extensively cited in recent years, a direct comparison to EMPS appears to be lacking in the literature.
We therefore briefly summarize the theory for blends before highlighting its similarities and differences with our approach for EMPS.

Let $\bm{x}_{\text{A}}$ and $\bm{x}_{\text{B}}$ denote the coordinates of points belonging to polymers A and B, respectively.
A ``polarization'' field $\bm{P}$ may then be defined as the displacement between their local centers of mass:
$\bm{P} = \langle \bm{x}_{\text{A}} \rangle - \langle \bm{x}_{\text{B}} \rangle$, where $\langle \cdot \rangle$ denotes the average over a small spatial region.
In the uniform phase, $\bm{P}=0$.
The scalar order parameter $\rho$, which describes local compositional fluctuations during phase separation, is identified with the effective
``charge'' density associated with $\bm{P}$. Extending the electrostatic analogy leads to Gauss' law
\begin{equation}
  \div\bm{P} = -\rho.
  \label{eq:polarization}
\end{equation}

The total elastic energy is taken to be the electrostatic self-energy associated with the bound charges.
It takes the form
\begin{equation}
  \mathscr{F}_{\text{el}} = \frac{1}{2}C \int\dd^{3}x\, \abs{\bm{P}}^{2},
  \label{eq:free_de_gennes_elastic}
\end{equation}
where the parameter $C$ is the strength of self-interaction.
Using Eq.~\eqref{eq:polarization}, we can write $\mathscr{F}_{\text{el}}$ entirely in terms of $\rho$
if only contributions from the longitudinal modes (i.e., those along $\bm{q}$) are considered.
Then, after including the usual Ginzburg--Landau terms [see Eq.~\eqref{eq:free_GL}], the total Gaussian (quadratic)
free energy in Fourier space becomes~\cite{gennes1979}
\begin{equation}
  \mathscr{F}_{\text{G}} = \frac{1}{2}\int\frac{\dd^{3}q}{(2\pi)^{3}}\left[a(T - \Tc) + \kappa q^{2} + \frac{C}{q^{2}}\right]\rho_{\bm{q}}\rho_{-\bm{q}}.
  \label{eq:free_de_gennes}
\end{equation}
Similar free energies have also been proposed to explain microphase separation in diblock copolymers~\cite{ohta1986}.
At sufficiently low temperature, such free energies result in the emergence of microphases owing to competition between the interfacial and elastic terms, $\kappa q^{2}$ and $Cq^{-2}$.
The scattering peak $\qm$ in the static structure factor associated with this microphase separation is $\qm^{4} = C/\kappa$.

De Gennes' phenomenological model is intended to describe the competition between interfacial effects whose strength
is controlled by $\kappa$ and elastic effects, which become relevant only at length scales close to $\xi$, the end-to-end distance between crosslinks.
Since $\kappa$ and $\xi$ provide the only relevant energy and length scale, dimensional analysis naturally leads to the choice $C = \kappa/\xi^{4}$.
As $\xi \sim N^{1/2}$ [Eq.~\eqref{eq:mesh_size_basic}] we see that the scattering peak $\qm$
scales with the number $N$ of monomers between the crosslinks as
\begin{equation}
  \qm = \left(C/\kappa\right)^{1/4} \sim \xi^{-1} \sim N^{-1/2}.
  \label{eq:degennes_qm}
\end{equation}
Similarly, one can show that the phase transition temperature $\Tm$ decreases linearly with $N^{-1}$.

Given that the underlying motivation of our approach is similar to that of de Gennes', we can compare the two as follows:

{\it i})~The Gauss law in Eq.~\eqref{eq:polarization} is analogous to the volume-conservation constraint in Eq.~\eqref{eq:matcons}.
Here, $\bm{P}$ plays a role similar to the displacement field $\bm{u}$, whereas $\rho$ corresponds to the uncoarse-grained order parameter $\psi$.
The key distinction lies in the coarse-graining procedure: in our approach, the order parameter is coarse-grained again,
whereas in Eq.~\eqref{eq:polarization} it is $\bm{P}$ that is coarse-grained (once, by construction).

{\it ii})~Our model employs the conventional elastic energy density, Eq.~\eqref{eq:free_el},
which is quadratic in the strain and, therefore, depends on gradients of the displacement field $\bm{u}$.
In contrast, Eq.~\eqref{eq:free_de_gennes_elastic} assumes an elastic energy that is quadratic in the displacement field ($\bm{P}$) itself.
The domain-size selection in Eq.~\eqref{eq:degennes_qm} is entirely a consequence of this particular choice of elastic energy.

{\it iii})~Both our model and Eq.~\eqref{eq:free_de_gennes} assume that the domain-size selection is governed solely by the longitudinal modes' contribution to the elastic energy.

{\it iv})~In both theories, the key length scale is $\xi$, the end-to-end distance between crosslinks.
It is this choice that gives rise to the experimentally observed scaling behavior in both theories.
In Eq.~\eqref{eq:free_de_gennes}, $\xi$ is only used to set the strength $C$ of the elastic self-interaction, whereas in our case,
it governs the length scale of nonlocal coarse-graining and is related to the elastic modulus [via Eq.~\eqref{eq:mesh_size}].

{\it v})~From Eq.~\eqref{eq:repeat_units}, we see that the dry Young's modulus follows the scaling $Y \sim N^{-1}$ with the number of monomers $N$.
Therefore, from Eqs.~\eqref{eq:qm_real} and \eqref{eq:Tm_real} of our model, we see that the EMPS phase transition temperature
$\Tm$ decreases linearly with $N^{-1}$ with the scattering peak $\qm \sim N^{-1/2}$, similar to blends.

{\it vi})~Even though the structure factor associated with Eq.~\eqref{eq:free_de_gennes} vanishes as $\bm{q}\to0$,
experiments on crosslinked blends~\cite{briber1988} have found a nonzero $S(\bm{q}\,{=}\,0)$, motivating alternative theoretical descriptions~\cite{read1995}.
In comparison, Eqs.~\eqref{eq:free_binary} and \eqref{eq:structure} from our model for EMPS predict a nonzero $S(\bm{q}\,{=}\,0)$.
This suggests that it can be extended to study blends as well.

\section{Concluding remarks}
\label{sec:conclusion}

We have presented a theoretical framework to explain elastic microphase separation (EMPS). Namely, the formation of microphases
of finite size in solvent-swollen elastomers~\cite{fernandez-rico2024}.
EMPS arises from a mismatch between the length scales governing elasticity and thermodynamics in elastomeric polymer networks.
Unlike previous approaches~\cite{qiang2024,paulin2026}, including our own~\cite{mannattil2025},
which relied on nonstandard theories of elasticity~\cite{eringen1987}, the current framework employs a nonlocal coupling between thermodynamics and makes use of conventional elasticity.
This approach provides an intuitive physical basis for EMPS, while remaining generalizable and compatible
with established theories of phase separation and density fluctuations in gels~\cite{onuki1993,onuki2002,dimitriyev2019}.

The predictions of the present framework can be directly compared with existing experimental results on EMPS~\cite{fernandez-rico2024}.
In particular, the theory captures the dependence of both the characteristic domain size and the phase transition temperature on elastomer stiffness.
For isotropically swollen elastomers, the predicted scaling relations are in good quantitative agreement with the experimentally reported trends~\cite{fernandez-rico2024}.
This agreement shows that network elasticity plays a central role in controlling the morphology and phase behavior of elastomers.

Our theory predicts that EMPS in isotropically swollen elastomers is a first-order phase transition,
consistent with known results for other systems exhibiting
microphase separation, such as block copolymers~\cite{fredrickson1987}.
The experimentally observed hysteresis-free and reversible nature of EMPS can be explained
by the vanishingly small phase-coexistence regions [see Fig.~\ref{fig:real_isotropic}].
Finally, as we have demonstrated, EMPS exhibits a scaling behavior similar to that observed
in crosslinked polymer blends~\cite{briber1988} and predicted earlier by de Gennes~\cite{gennes1979}.
The phase behavior of crosslinked polymer blends and gels can be drastically altered by anisotropic effects~\cite{onuki2002}.
Extending the present framework to account for anisotropy in elastomers is therefore a natural next step and will be the focus of Part~II~\cite{part2}.

Elastomers exhibit a wide range of complex behaviors.
Other natural extensions to the questions we have considered in this paper include the phase-separation kinetics, extensions to ternary and multicomponent systems, volume phase transitions, etc.~\cite{onuki2002}.
Our framework may also be applied to similar systems where elastic and thermodynamic effects are often governed by distinct length scales, such as certain porous materials~\cite{coussy2004,derr2020,paulin2022} and colloidal suspensions~\cite{tanaka2005}.

\subsection*{Acknowledgements}

M.M.~thanks L.~Mahadevan and Mehrana Nejad for fruitful discussions.
H.D.~acknowledges support from the Israel Science Foundation (ISF Grant No.~1611/24).
D.A.~acknowledges support from the Israel Science Foundation (ISF Grant No.~226/24).

\bibliography{library,misc}

\end{document}